\documentclass[superscriptaddress,amsmath,amssymb,prd,preprintnumbers,showpacs,twocolumn,nofootinbib]{revtex4-2}

\usepackage{hyperref}
\usepackage{amsmath}
\usepackage{graphicx}
\usepackage{epsf}
\usepackage{psfrag}
\usepackage{epsfig}
\usepackage{graphics}
\usepackage{amsfonts}
\usepackage{epstopdf}
\usepackage{slashed}
\usepackage{footnote}
\usepackage[utf8]{inputenc}
\usepackage[T1]{fontenc}
\usepackage[makeroom]{cancel}
\usepackage{dsfont}
\usepackage{subfigure}
\usepackage{xcolor}
\usepackage{units}
\usepackage{upgreek}
\usepackage{booktabs}

\begin{document}

\newcommand{\be}{\begin{equation}}
\newcommand{\ee}{\end{equation}}
\newcommand{\bq}{\begin{eqnarray}}
\newcommand{\eq}{\end{eqnarray}}
\newcommand{\ba}{\begin{align}}
\newcommand{\ea}{\end{align}}

\newcommand{\Dslash}{\hbox{$\partial\!\!\!{\slash}$}}
\newcommand{\qslash}{\hbox{$q\!\!\!{\slash}$}}
\newcommand{\pslash}{\hbox{$p\!\!\!{\slash}$}}
\newcommand{\bslash}{\hbox{$b\!\!\!{\slash}$}}
\newcommand{\kslash}{\hbox{$k\!\!\!{\slash}$}}
\newcommand{\kbruto}{\hbox{$k \!\!\!{\slash}$}}
\newcommand{\pbruto}{\hbox{$p \!\!\!{\slash}$}}
\newcommand{\qbruto}{\hbox{$q \!\!\!{\slash}$}}
\newcommand{\lbruto}{\hbox{$l \!\!\!{\slash}$}}
\newcommand{\bbruto}{\hbox{$b \!\!\!{\slash}$}}
\newcommand{\parbruto}{\hbox{$\partial \!\!\!{\slash}$}}
\newcommand{\Abruto}{\hbox{$A \!\!\!{\slash}$}}
\newcommand{\bbbruto}{\hbox{$b_1 \!\!\!{\slash}$}}
\newcommand{\bbbbruto}{\hbox{$b_2 \!\!\!{\slash}$}}

\newcommand{\orcid}[1]{\href{https://orcid.org/#1}{\includegraphics[width=10pt]{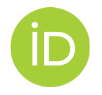}}}

\title{Vacuum Cherenkov radiation for nonminimal dimension-5 Lorentz violation}

\author{A.Yu.~Petrov\orcid{0000-0003-4516-655X}} 
\email{petrov@fisica.ufpb.br}
\affiliation{Departamento de F\'isica, Universidade Federal da Para\'iba, Caixa Postal 5008, 58051-970, Jo\~ao Pessoa (PB), Brazil}

\author{M.~Schreck\orcid{0000-0001-6585-4144}}
\email{marco.schreck@ufma.br}
\affiliation{Departamento de F\'isica, Universidade Federal do Maranh\~ao, 65080-805 S\~ao Lu\'is (MA), Brazil}

\author{A.R.~Vieira\orcid{0000-0002-5583-1560}}
\email{alexandre.vieira@uftm.edu.br}
\affiliation{Instituto de Ciências Agrárias, Exatas e Biológicas de Iturama -- ICAEBI, Universidade Federal do Triângulo Mineiro --
UFTM, Iturama, 38280-000, Minas Gerais, Brazil}

\date{\today}

\begin{abstract}
Vacuum Cherenkov radiation is investigated in the Lorentz-violating Standard-Model Extension for isotropic dim-5 operators $\hat{m}$ and $\hat{a}^{\mu}$ in the fermion sector. Both the kinematics and dynamics of this process are analyzed by analytical and numerical means, leading to its decay and radiated-energy rates as functions of the initial-fermion momentum. We adopt the point of view that vacuum Cherenkov radiation is actually a physical phenomenon expected to occur for a charged, massive fermion in the presence of Lorentz violation, when some additional requirements are satisfied. The absence of this effect in ultrahigh-energy cosmic rays detected on Earth allows us to infer stringent bounds on isotropic dim-5 Lorentz violation in protons, quarks, and electrons.
\end{abstract}

\pacs{11.30.Cp, 03.65.Pm, 03.70.+k, 95.85.Ry}
\keywords{Lorentz violation; Dirac equation; Quantum field theory; Cosmic rays}

\maketitle

\section{Introduction}
\label{s0}

Lorentz and CPT symmetries are known as the most important ingredients in the construction of any relativistic field theory model. Dimensional arguments hint towards a unification of the Standard Model and General Relativity at the Planck scale $E_{\mathrm{Pl}}\sim \unit[10^{19}]{GeV}$. Therefore, it is expected that the fundamental spacetime symmetries are spontaneously broken at energies of this magnitude, which was, indeed, demonstrated to occur in certain string field theories~\cite{Kostelecky:1988zi,Kostelecky:1989jp,Kostelecky:1989jw,Kostelecky:1991ak,Kostelecky:1994rn}. Physical mechanisms for spacetime symmetry breaking were also identified in loop quantum gravity~\cite{Gambini:1998it,Bojowald:2004bb}, noncommutative spacetimes~\cite{Carroll:2001ws}, spacetime foam models~\cite{Klinkhamer:2003ec,Bernadotte:2006ya}, and even settings of chiral field theories on topologically nontrivial spacetime manifolds~\cite{Klinkhamer:1999zh,Klinkhamer:2002mj,Ghosh:2017iat}. If Lorentz invariance is, indeed, fundamentally broken at $E_{\mathrm{Pl}}$, it may be possible to see tiny remnant effects even for $E\ll E_{\mathrm{Pl}}$.

The field theory framework known as the Lorentz-violating (LV) Standard Model Extension (SME)~\cite{Colladay:1996iz,ColKost} modifies the action of the Standard Model as well as the Einstein-Hilbert action of gravity at an effective level by adding terms that violate (local) Lorentz invariance or CPT invariance or both. In the Standard-Model particle sectors, these contributions break the spacetime symmetries explicitly, since they incorporate background fields that do not arise dynamically. Contrarily, explicit breaking in gravity must be handled with great care \cite{Kostelecky:2003fs,Bluhm:2014oua,Bluhm:2015dna,Bluhm:2016dzm,Bluhm:2019ato,Kostelecky:2020hbb,Bailey:2024zgr,Reyes:2024ywe}, 
whereupon most dedicated studies of such settings rely on spontaneous spacetime symmetry violation.

Any experimental signal of a LV effect would support the idea of a fundamental unified theory at the Planck scale that spontaneously breaks Lorentz symmetry via a physical mechanism. However, even if someone has the viewpoint that Lorentz and CPT symmetries are exact in nature, an important question to ask is what their applicability limits are based on state-of-the-art experiments. After all, physical principles and theories should always be subject to experimental tests. The SME serves as a comprehensive framework for performing these tests and to quantify the extend to which Lorentz and CPT invariance are guaranteed. Since 2008, such experimental results have been compiled in yearly updated data tables~\cite{DataTables}. A general review of various aspects of LV theories, including a discussion of experimental searches for Lorentz violation, is presented in Ref.~\cite{bookLV}.

The present work is focused on the nongravitational LV SME~\cite{ColKost}. On the one hand, all renormalizable Lorentz and CPT-violating contributions with operators of mass dimensions $d\leq 4$ compose the minimal LV SME. When it comes to the latter, Lorentz invariance has been largely tested in the matter and the Abelian gauge sector; see Ref.~\cite{DataTables}. On the other hand, the nonminimal SME includes operators with mass dimensions $d>4$. Nonrenormalizable operators of $d=5,6$ in photons, electrons, muons, quarks, and neutrinos have already been constrained significantly. These bounds partially extend to operators of up to $d=10$.

Our interest is in a particular unusual particle-physics process that plays a significant role in the presence of Lorentz violation and is known as vacuum Cherenkov radiation. Recall that Cherenkov radiation in optical media occurs when a massive, charged particle travels through the medium with a velocity greater than the phase velocity of light in this medium \cite{Jelley:1958}. At the same time, Lorentz and CPT symmetry violations are known to provide several modifications of the usual physics, including anisotropic dispersion relations, vacuum birefringence, and the rotation of the polarization plane of electromagnetic waves~\cite{KosMew,Alan}. Lorentz violation in photons is effectively described by a vacuum with a nontrivial refractive index. The latter enables Cherenkov-type emissions of photons by charged particles \textit{in vacuo}~\cite{Beall:1970rw,Coleman:1997xq,Ralf,Ralf2,Lehnert:2004bjc}.

There are two different points of view when it comes to vacuum Cherenkov radiation. The first interprets this process as an effect caused by instabilities of the LV theory due to negative-energy states in nonconcordant frames~\cite{Alan}. These instabilities could be eliminated if higher-order operators were taken into account. According to very recent studies \cite{AlanKostelecky:2024gek,AlanKostelecky:2024psy}, instabilities in certain field theory models are avoided completely, when the vacuum state is physically determined by coupling the model to a heat bath. The latter is possibly formed from known matter and radiation in the Universe, where dark matter and dark energy can play a role, as well.

In light of these findings, the second point of view takes the stability of the modified theory seriously. Thereupon, vacuum Cherenkov radiation is an observable phenomenon, in analogy with Cherenkov radiation in an optical medium. As long as this process is not observed, data from ultrahigh-energy cosmic rays (UHECRs) can be used to put stringent bounds on CPT- and Lorentz-violating coefficients of the SME. In this work, we adopt the second attitude and employ such data to bound certain nonminimal dim-5 LV coefficients in the fermion sector. Thus, nonminimal operators are not added to fix an instability of the minimal sector, but their physical implications are explored independently of the minimal SME.

So far, vacuum Cherenkov radiation has been widely investigated in the SME photon sector, which decomposes into a CPT-even piece, parametrized by the fourth-rank tensor-valued background field $k_F$, and the CPT-odd Carroll-Field-Jackiw (CFJ) term governed by the vector-valued background field $k_{AF}$. In Refs.~\cite{Ralf,Ralf2} the process was studied at the classical level in Maxwell-CFJ (MCFJ) theory with purely spacelike $k_{AF}$, while these findings are complemented in Refs.~\cite{Klinkhamer,Klinkhamer2} by taking into account quantum effects. References~\cite{Schober,Brett2} delve into the classical modified electrodynamics based on purely timelike MCFJ theory, where vacuum Cherenkov radiation was demonstrated not to occur.

Quantum aspects such as microcausality and unitarity in MCFJ theory are on the menu of Ref.~\cite{Klinkhamer4}, with issues shown to arise for a purely timelike $k_{AF}$. For a consistent quantization, timelike MCFJ theory requires a photon mass \cite{Colladay,Colladay:2016eaw,Ferreira:2020wde}. In fact, the full quantum treatment of vacuum Cherenkov radiation revealed that it has a nonzero decay rate, whose low-energy regime is suppressed quadratically by Lorentz violation~\cite{Colladay}. The findings of the latter work were extended by considering a general class of covariant gauges~\cite{Colladay2}. MCFJ theory was also coupled to other particles like pions~\cite{Brett3} to shed light on Cherenkov-type radiation of pions by photons in the presence of a CPT-violating term.

Studies of Cherenkov-type processes in the CPT-even extension of electrodynamics are more scarce, but a certain interest in such scenarios has also arisen~\cite{Brett,Ralf3,Klinkhamer3,Klinkhamer5,Klinkhamer:2007ak,Klinkhamer:2008ss,Klinkhamer:2008dw,Klinkhamer:2017puj,Duenkel:2021gkq,Risse:2022unt,Duenkel:2023nlk}. Most of the latter papers are dedicated to the isotropic sector of the $k_F$ term governed by the single coefficient~$\tilde{\kappa}_{\mathrm{tr}}$, which has been constrained from collider and astrophysics data, respectively. The analyses of Refs.~\cite{Ralf3,Klinkhamer3} are comparable from a theoretical viewpoint, where Ref.~\cite{Klinkhamer5} is an extension of Ref.~\cite{Klinkhamer3} based on the parton model.

Our understanding of vacuum Cherenkov radiation has also improved in settings different from the SME. Such research has mainly been based on modified dispersion relations motivated by quantum gravity. Cases of higher-power momenta are studied in Refs.~\cite{Jacobson:2002hd,Jacobson:2005bg,Carmona:2014lqa,Martinez-Huerta:2016odc,Martinez-Huerta:2017unu,Martinez-Huerta:2017ntv,Saveliev:2024whq,Li:2025uwn}. The authors of Refs.~\cite{Kalaydzhyan:2015ija,Kalaydzhyan:2016lfo} look at purely minimal alterations caused by the gravitational potential of the Earth. Additionally, generic field theory approaches are explored, e.g., in Refs.~\cite{Maccione:2007yc,Anselmi:2011ae,Rubtsov:2012kb}. Specific field theory models based on noncommutative spacetimes \cite{Castorina:2004hv}, nonlinear electrodynamics \cite{Macleod:2018zcb}, Lifshitz electrodynamics \cite{Bufalo:2021myy}, and modified electrodynamics originating from a gravitational background with a scalar field \cite{vandeBruck:2016cnh} are investigated, too.

There has also been interest in Cherenkov-type emission of gravitons in modified linearized gravity~\cite{quarkfrac,Jay2,Marco2,Artola:2024mky}. The latter is a gauge theory, which renders such analyses very similar to those for modified photons. Last but not least, the development of theoretical tools for a proper treatment of LV processes in fermions at tree-level \cite{Marco3} enabled computations of decay rates of vacuum Cherenkov radiation for a subset of coefficients of the minimal SME fermion sector~\cite{Marco,Schreck:2017egi}; see also Refs.~\cite{Borisov:2023xus,Borisov:2024gzb} for additional studies not relying on these techniques. Note that any investigation with modified fermions is more involved, since a fermion occurs as both an initial and a final particle in this process.

Although Cherenkov-type radiation \textit{in vacuo} has been subject to broad studies within the minimal SME, understanding the properties of this process in the presence of nonminimal Lorentz violation in fermions and photons has largely remained an open problem. This is mainly due to technical reasons. Many nonminimal coefficients give rise to complicated dispersion relations, including spurious ones~\cite{Marco3}, which render most analytical computations challenging. The present work is intended to fill this gap for certain isotropic nonminimal dim-5 coefficients in Dirac fermions. Unlike computations for the minimal sector, which can often be carried out analytically at all orders in Lorentz violation, here we have to restrict ourselves to leading order. Numerical approaches will be indispensable to compute decay and radiated-energy rates at all orders in Lorentz violation.

The structure of the paper is as follows. The purpose of Sec.~\ref{s1} is to review the basic properties of modified Dirac fermions in the SME and to define the necessary quantities. Generic formulas for decay and radiated-energy rates within isotropic settings are set up in Sec.~\ref{s2}. Sections~\ref{s3} and \ref{s4} present results for the decay rates of the isotropic dim-5 $m$ and $a$ coefficients, respectively. The asymptotic behaviors of the radiated-energy rates are looked at more closely in Sec.~\ref{sec:radiated-energy-rate}. Here, the goal is to understand how quickly Dirac fermions lose energy when they are subject to vacuum Cherenkov radiation. In Sec.~\ref{s5}, the previous findings culminate in a number of constraints on isotropic dim-5 LV in protons, quarks, and electrons. Last but not least, the results are discussed and concluded in Sec.~\ref{s6}. Natural units with $\hbar=c=1$ are used, unless otherwise stated.

\section{Vacuum Cherenkov process}
\label{s1}

Our starting point is a modified quantum electrodynamics (QED) with the usual photon sector but Dirac fermions modified by operators of the nonminimal SME:
\begin{subequations}
\label{eq:modified-QED}
\begin{align}
\mathcal{L}_{\mathrm{QED}}&=\mathcal{L}_{\upgamma}+\mathcal{L}_{\psi}\,, \\[1ex]
\mathcal{L}_{\upgamma}&=-\frac{1}{4}F^{\mu\nu}F_{\mu\nu}-\frac{1}{2}(\partial_{\mu}A^{\mu})^2\,, \\[1ex]
\mathcal{L}_{\psi}&=\frac{1}{2}\overline{\psi}(\gamma^{\mu}i\mathcal{D}_{\mu}-m_{\psi}+\widehat{\mathcal{Q}})\psi+\text{H.c.}\,,
\end{align}
\end{subequations}
with \textit{U}(1) gauge field $A_{\mu}$, electromagnetic field strength tensor $F^{\mu\nu}=\partial^{\mu}A^{\nu}-\partial^{\nu}A^{\mu}$, Dirac spinor field $\psi$, Dirac conjugate spinor field $\overline{\psi}:=\psi^{\dagger}\gamma^0$, and fermion mass $m_{\psi}$. All fields are defined in Minkowski spacetime with the metric tensor $\eta_{\mu\nu}$ of signature $(+,-,-,-)$. The Dirac matrices $\gamma^{\mu}$ satisfy the Clifford algebra $\{\gamma^{\mu},\gamma^{\nu}\}=2\eta^{\mu\nu}$. The photon sector is minimally coupled to the fermion sector via the gauge-covariant derivative $\mathcal{D}_{\mu}=\partial_{\mu}+\mathrm{i}qA_{\mu}$ with the electric charge $q$ of the fermion. The operator $\widehat{\mathcal{Q}}$ contains all nonminimal LV coefficients compatible with coordinate invariance and gauge symmetry. We are interested in the isotropic pieces of the nonminimal dim-5 $\hat{m}$ and $\hat{a}^{\mu}$ operators, which are specifically described by
\begin{subequations}
\begin{equation}
\label{eq:Qhat-m}
\widehat{\mathcal{Q}}=(m^{(5)})^{\alpha\beta}\partial_{\alpha}\partial_{\beta}\mathds{1}_4\,,
\end{equation}
and
\begin{equation}
\label{eq:Qhat-a}
\widehat{\mathcal{Q}}=(a^{(5)})^{\mu\alpha\beta}\gamma_{\mu}\partial_{\alpha}\partial_{\beta}\,,
\end{equation}
\end{subequations}
respectively, where $(m^{(5)})^{\alpha\beta}$ and $(a^{(5)})^{\mu\alpha\beta}$ are the corresponding controlling coefficients. Furthermore, $\mathds{1}_4$ is the identity matrix in spinor space. For definiteness, we will be working in the Dirac representation.

\begin{figure}
\centering
\includegraphics{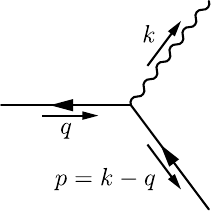}
\caption{Tree-level Feynman diagram for vacuum Cherenkov radiation of a positively charged Dirac fermion. The incoming fermion carries four-momentum $q$, whereas the outgoing photon and fermion carry four-momentum $k$ and $p=q-k$, respectively.} 
\label{fig0}
\end{figure}

The nonminimal coefficients chosen are spin-degenerate. This means that there is no difference in the dispersion relations for spin-up and spin-down fermions. Therefore, spin-flip processes, which play a role for spin-nondegenerate coefficients~\cite{Marco,Schreck:2017egi}, are insignificant. For a process to be allowed energetically, energy-momentum must be conserved. Thus, we define the energy balance by
\begin{equation}
\Delta E\equiv E({\bf q})-|{\bf k}|-E({\bf q}-{\bf k})\,,
\label{ebeq}
\end{equation} 
where $\mathbf{q}$ is the spatial momentum of the initial fermion and $\mathbf{k}$ the momentum of the emitted photon. The fermion dispersion relation as a function of the generic momentum $\mathbf{p}$ is denoted as $E(\mathbf{p})$. 
Energy-momentum conservation demands that $\Delta E=0$. For the kinematics we must take into account that the fermion dispersion relations are affected by Lorentz violation. The latter are obtained by setting the determinant of the modified Dirac operator to zero and solving for $p^0$. The decay rate is calculated from the single tree-level diagram presented in Fig.~\ref{fig0}. As mentioned above, the computation of the decay rate for fermions is feasible courtesy of a set of tools developed in previous works~\cite{Marco,Marco3}.

The amplitude corresponding to the Feynman diagram in Fig.~\ref{fig0} follows from a set of Feynman rules for the external lines. External fermion lines are described by modified spinor solutions $u^{(s)}(q)$ of the Dirac equation in momentum space, whereas external photon lines stand for the standard electromagnetic polarization vectors $\varepsilon^{\mu}(k)$. Besides the usual Feynman rule for the QED vertex, another one arises for SME operators that are contracted with additional spacetime derivatives. The fermion propagator is also modified. While the latter is not needed to compute the tree-level diagram of Fig.~\ref{fig0}, it can be used to derive altered completeness relations for the spinor solutions of the Dirac equation; see Ref.~\cite{Marco3} for minimal modifications. However, in the present work, we will compute these expressions directly from the spinors.

Due to \textit{U}(1) gauge invariance, the QED Ward identity still holds for on-shell photons, even when the fermions are subject to Lorentz violation. To see this, we consult the modified Dirac equation for the dim-5 operators $\hat{m}$ and $\hat{a}^{\mu}$:
\begin{subequations}
\label{eq:modified-dirac}
\begin{equation}
0=(p^{\alpha}\gamma_{\alpha}-m_{\psi}+\widehat{\mathcal{Q}})\psi=(p^{\alpha}\gamma_{\alpha}-\widehat{M}(p))\psi\,,
\end{equation}
with
\begin{equation}
\label{eq:Mhat-m}
\widehat{M}(p)=(m_{\psi}+(m^{(5)})^{\alpha\beta}p_{\alpha}p_{\beta})\mathds{1}_4\,,
\end{equation}
and
\begin{equation}
\label{eq:Mhat-a}
\widehat{M}(p)=m_{\psi}\mathds{1}_4+(a^{(5)})^{\mu\alpha\beta}\gamma_{\mu}p_{\alpha}p_{\beta}\,,
\end{equation}
\end{subequations}
respectively. In general, we start from the relationship
\begin{align}
k_{\mu}\gamma^{\mu}+\widehat{M}(p_2)-\widehat{M}(p_1)&=(p_{1,\mu}\gamma^{\mu}-\widehat{M}(p_1)) \notag \\
&\phantom{{}={}}-(p_{2,\mu}\gamma^{\mu}-\widehat{M}(p_2))\,,
\end{align}
with the difference $k\equiv p_1-p_2$ of the four-momenta $p_1,p_2$. Sandwiching both sides of the latter with $\bar{u}(p_2)$ from the left and $u(p_1)$ from the right and using the modified Dirac equation~\eqref{eq:modified-dirac} implies
\begin{equation}
0=\overline{u}(p_2)\Big[k_{\mu}\gamma^{\mu}+\widehat{M}(p_2)-\widehat{M}(p_1)\Big]u(p_1)\,.
\end{equation}
By inserting the explicit forms of $\widehat{M}(p)$ for the dim-5 operators $\hat{m}$ and $\hat{a}^{\mu}$, we arrive at the Ward identity
\begin{subequations}
\label{eqWI}
\begin{equation}
k_{\mu}\mathcal{M}^{\mu}=0\,,
\end{equation}
with the expressions
\begin{align}
\mathcal{M}^{\mu}&=\overline{u}(p_2)\Big[\gamma^{\mu}-(m^{(5)})^{\mu\nu}(p_1+p_2)_{\nu}\mathds{1}_4\Big]u(p_1)\,, \\[2ex]
\mathcal{M}^{\mu}&=\overline{u}(p_2)\Big[\gamma^{\mu}-(a^{(5)})^{\nu\mu\varrho}\gamma_{\nu}(p_1+p_2)_{\varrho}\Big]u(p_1)\,,
\end{align}
\end{subequations}
for $m^{(5)}_{\alpha\beta}$ and $a^{(5)}_{\alpha\beta\gamma}$, respectively.

Nonminimal operators, just as some minimal ones, can introduce additional time derivatives into the theory that lead to a nonconventional time evolution of the asymptotic fermion states~\cite{crosssec}. This is of certain importance for us, since the tree-level amplitude of the process involves such states in any case. For minimal operators, additional time derivatives only occur in $\Gamma^0$. Here, $\Gamma^{\mu}$ is defined in Ref.~\cite{ColKost} and can be understood as a ``generalized'' set of Dirac matrices, which is conveniently employed to cast the SME fermion action into a form reminiscent of the Dirac action. Issues with additional time derivatives are avoided via the spinor field redefinition $\psi=A\chi$, where $A$ is a $(4\times 4)$ matrix that obeys the relation $A^{\dagger}\gamma^0\Gamma^0A=\mathds{1}_4$. However, the situation is different for nonminimal operators. Then, additional time derivatives are not linked to the specific component $\Gamma^0$, but are always present in certain observer frames. They can even occur for operators whose minimal counterparts are not accompanied by any time derivatives, such as $\hat{a}^{\mu}$.

One possibility of tackling this problem is to replace each $p^0$ in the LV terms by the standard massive dispersion relation~\cite{Schreck:2014qka}. However, this technique only works at first order in Lorentz violation, whereas we intend to perform at least some calculations at higher orders or even exact in Lorentz violation. Having said this, we will proceed without eliminating the additional time derivatives and judge from the physical results whether or not this procedure can be deemed reasonable in the cases to be considered.

We are now ready to build the amplitude $\mathcal{M}$ and its square for the Cherenkov process presented in Fig.~\ref{fig0}. It is given by
\begin{subequations}
\begin{align}
\mathcal{M}&=e\bar{u}^{(s)}(q-k)\Gamma^{\mu} u^{(s')}(q)\epsilon_{\mu}^{(\lambda)}(k)\,, \\[1ex]
|\mathcal{M}|^2&=e^2\bar{u}^{(s)}_{a}(q-k)\Gamma^{\mu}_{ab} u^{(s')}_{b}(q)\epsilon^{(\lambda)}_{\mu}(k)
\bar{u}^{(s)}_{c}(q-k) \nonumber \\
&\phantom{{}={}}\times \Gamma^{\nu}_{cd} u^{(s')}_{d}(q)\epsilon^{(\lambda)}_{\nu}(k)\,,
\label{eqamp}
\end{align}
\end{subequations}
where $s,s'$ are spin labels and $\lambda$ is a polarization label. The remaining Latin indices identify components in spinor space. Note that the modified spinor solutions of the Dirac equation~\eqref{eq:modified-dirac} must be employed here. Although we do not make use of the object $\Gamma^{\mu}$ in the original Lagrange density of Eq.~\eqref{eq:modified-QED}, we introduce an analogous set $\Gamma^{\mu}$ of $(4\times 4)$ matrizes in spinor space at the level of the modified $\overline{\psi}\psi A^{\mu}$ vertex. In principle, the needed Feynman rule can be inferred from the Ward identity of Eq.~\eqref{eqWI}. In particular,
\begin{subequations}
\begin{equation}
\begin{array}{c}
\vspace{-0.1cm}
\includegraphics[scale=0.75]{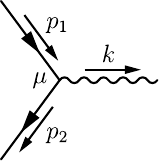}
\end{array}=\mathrm{i}e\Gamma^{\mu}\,,
\end{equation}
with the elementary charge $e$, where
\begin{align}
\Gamma^{\mu}&=\gamma^{\mu}-(m^{(5)})^{\mu\nu}(p_1+p_2)_{\nu}\mathds{1}_4\,, \\[2ex]
\Gamma^{\mu}&=\gamma^{\mu}-(a^{(5)})^{\varrho\mu\nu}\gamma_{\varrho}(p_1+p_2)_{\nu}\,,
\end{align}
\end{subequations}
for the dim-5 operators $\hat{m}$ and $\hat{a}^{\mu}$, respectively.

Moreover, the initial fermions are unpolarized, which is why we average over the initial polarizations~\cite{Peskin:1995}. Due to degeneracy of the photon polarization states and the fermion spin states, all possible final polarizations and spins must be summed over. Then,
\begin{subequations}
\begin{align}
\bar{|\mathcal{M}|^2}&=\frac{1}{2}\sum_{\text{pol}} |\mathcal{M}^{(s,s')}|^2 \notag \\
&=2\pi \alpha \sum_{s,s'} u^{(s')}_{d}(q) \bar{u}^{(s')}_{a}(q)\Gamma^{\mu}_{ab}u^{(s)}_{b}(q-k) \nonumber \\
&\phantom{{}={}}\times \bar{u}^{(s)}_{c}(q-k)\Gamma^{\nu}_{cd}\sum_{\lambda}\epsilon^{(\lambda)}_{\mu}(k)\epsilon^{(\lambda)}_{\nu}(k) \nonumber\\
&=2\pi \alpha \sum_{s,s'}\mathrm{Tr}[\Lambda^{(s')}(q)\Gamma^{\mu}\Lambda^{(s)}(q-k)\Gamma^{\nu}]\Pi_{\mu\nu}(k)\,,
\label{eqamps}
\end{align}
with the fine-structure constant $\alpha=e^2/(4\pi)$, where
\begin{align}
\label{eq:spinor-bilinears}
\Lambda^{(s)}_{ab}(q)&=u^{(s)}_a(q)\bar{u}^{(s)}_b(q)\,, \\[1ex]
\Pi_{\mu\nu}(k)&=\sum\limits_{\lambda}\epsilon^{(\lambda)}_{\mu}(k)\epsilon^{(\lambda)}_{\nu}(k)\,,
\end{align}
\end{subequations}
and the trace is computed over the spinor indices.
For the $\hat{m}$ operator, the sum over the spinor bilinears is obtained directly from the $u$-type spinor solutions of the modified Dirac equation~\eqref{eq:modified-dirac}, which are explicitly stated in App.~\ref{apA}. It is found to be of the form
\begin{equation}
\label{eq:sum-spinor-projetors-m}
\sum_s \Lambda^{(s)}_{ab}(q)=[q^{\alpha}\gamma_{\alpha}+\widehat{M}(q)]_{ab}|_{q^0=E(\mathbf{q})}\,,
\end{equation}
in terns of $\widehat{M}$ given by Eq.~\eqref{eq:Mhat-m} and the suitable modified dispersion relation $E(\mathbf{q})$. The analogous result for $\hat{a}^{\mu}$ requires additional ingredients and will only be given later, at the appropriate time. The photon polarization tensor is decomposed as
\begin{equation}
\Pi^{\mu\nu}(k)=-\eta^{\mu\nu}-\frac{1}{{\bf k}^2}k^{\mu}k^{\nu}+\frac{1}{{|\bf k}|}(k^{\mu}n^{\nu}+k^{\nu}n^{\mu})\,,
\end{equation}
where $n^{\mu}$ is an auxiliary four-vector included to cast the polarization tensor into a covariant form. Each term in $\Pi^{\mu\nu}$ depending on a four-momentum with a free Lorentz index can be disregarded when $\Pi^{\mu\nu}$ is contracted with a gauge-invariant expression. Here, we benefit from the Ward identity of Eq.~\eqref{eqWI} figured out previously. Thus, $n^{\mu}$ is deduced to not have any physical meaning.

\section{Decay and radiated-energy rates for isotropic frameworks}
\label{s2}

In this section, we set up generic formulas for the decay and radiated-energy rates that describe the dynamics of vacuum Cherenkov radiation. The latter hold for sets of isotropic SME coefficients and pose the theoretical foundation for the physical results of interest in the forthcoming sections. It is convenient to express the space phase for the vacuum Cherenkov process in terms of the photon momentum $k^{\mu}$ in spherical coordinates $(k,\theta, \phi)$, where $k\equiv|{\bf k}|$ is the norm of the spatial photon momentum, $\theta$ the polar angle, and $\phi$ the azimuthal one. The spatial photon momentum points along the radial direction, and the polarization vectors
are perpendicular to it:
\begin{subequations}
\begin{align}
k^{\mu}&=k(1,\hat{\mathbf{k}})\,,\quad \hat{\mathbf{k}}=\begin{pmatrix}
\sin \theta \cos \phi \\
\sin \theta \sin \phi \\
\cos \theta \\
\end{pmatrix}\,, \\[1ex]
\epsilon^{(1)\mu}&=\begin{pmatrix}
0 \\
\cos \theta\cos \phi \\
\cos \theta\sin \phi \\
-\sin\theta \\
\end{pmatrix}\,, \quad\epsilon^{(2)\mu}=\begin{pmatrix}
0 \\
-\sin \phi \\
\cos \phi \\
0 \\
\end{pmatrix}\,.
\end{align}
\end{subequations}
The decay rate results from integrating the amplitude squared over the space phase of the outgoing particles:
\begin{subequations}
\label{eq:decay-rate}
\begin{align}
\Gamma&=\frac{1}{2E({\bf q})}\gamma\,, \\[1ex]
\gamma&=\int\frac{\mathrm{d}^3k}{(2\pi)^3}\int \frac{\mathrm{d}^3p}{(2\pi)^3}\frac{(2\pi)^4\delta^{(4)}(q-k-p)}{4\omega({\bf k})E({\bf p})}|\mathcal{M}|^2\,,
\label{eq2.1a}
\end{align}
\end{subequations}
with the decay constant $\gamma$ and the photon dispersion relation $\omega({\bf k})\equiv |{\bf k}|$. We can directly integrate over $\phi$ for isotropic cases. In terms of spherical coordinates, Eq.~(\ref{eq2.1a}) is rewritten as
\begin{equation}
\gamma=\frac{1}{8\pi}\int^{\infty}_0 \mathrm{d}k\, k^2\int^{\pi}_{0}\mathrm{d}\theta \frac{\sin \theta}{k E({\bf q}-{\bf k})}\delta(\Delta E)|\mathcal{M}|^2\,.
\end{equation}
The range of the photon momentum is restricted by the solutions for $\theta$. Finding zeros of the energy balance~\eqref{ebeq} leads to solutions $\theta=\theta_0=\theta_0(q,k,X_{\subset})$, where $-1\leq \cos \theta_0\leq 1$ and $X_{\subset}$ is a subset of LV coefficients. After computing the integral over $\theta$ and evaluating the $\delta$ function, the decay constant for isotropic coefficients is cast into the form
\begin{subequations}
\label{eq0.2}
\begin{align}
\gamma&=\frac{1}{8\pi}\int_0^{k_{\mathrm{max}}}\mathrm{d}k\, \Pi(k)|\mathcal{M}|^2|_{\theta=\theta_0}\,, \\[1ex]
\Pi(k)&=\frac{k\sin \theta}{E({\bf q}-{\bf k})}\left|\frac{\partial \Delta E}{\partial \theta}\right|^{-1}\Big|_{\theta=\theta_0}\,,
\end{align}
\end{subequations}
where $k_{\mathrm{max}}$ is the maximum possible value for the photon momentum, restricted by the requirement that $-1\leq \cos \theta_0\leq 1$.

A minor adaption of the previous integrand leads to the radiated-energy rate, which describes the energy emitted by the fermion in an infinitesimal time interval \cite{Klinkhamer2}:
\begin{equation}
\label{eq:radiated-energy-rate}
\frac{\mathrm{d}W}{\mathrm{d}t}=\frac{1}{8\pi}\int_0^{k_{\mathrm{max}}}\mathrm{d}k\, \omega\Pi(k)|\mathcal{M}|^2|_{\theta=\theta_0}\,.
\end{equation}
The integrals in Eqs.~\eqref{eq0.2}, \eqref{eq:radiated-energy-rate} are challenging to solve analytically for the examples that we are going to present below. Therefore, they will be evaluated with numerical tools provided by {\it Mathematica}. Even if it were possible to state the decay rates in a closed form, the expressions would not be illuminating. Hence, we are going to present plots of these functions. At least, it has turned out to be possible to obtain asymptotic behaviors by analytical means. Explicitly, we will be exploring the LV operators $\hat{m}$ and $\hat{a}^{\mu}$ involving subsets of the coefficients $m^{(5)}_{\alpha\beta} $ and $a^{(5)}_{\alpha\beta\gamma}$, respectively; see Refs.~\cite{Mewes,Zonghao}. Note that these operators have been studied earlier within the perturbative framework; see Refs.~\cite{TM1,TM2}. However, their impact on vacuum Cherenkov radiation has never been considered until now.

\section{Dimension-5 CPT-even operator}
\label{s3}

The dim-5 operator $\hat{m}$ contains the CPT-even coefficients $m^{(5)}_{\alpha\beta}$, which give rise to the term
\begin{equation}
-\frac{1}{2}(m^{(5)})^{\alpha\beta}\overline{\psi}\mathrm{i}D_{(\alpha}\mathrm{i}D_{\beta)}\psi+\text{H.c.} \subset \mathcal{L}^{(5)}_{\psi D}\,,
\end{equation}
of Tab.~I in Ref.~\cite{Zonghao}. Since the background tensor is not necessarily traceless, it possesses two independent isotropic pieces denoted as $\mathring{m}_{0,2}$. The nonminimal LV operator of Eq.~\eqref{eq:Qhat-m} in momentum space is given by
\begin{equation}
\widehat{\mathcal{Q}}=-m^{(5)}_{\alpha\beta}p^{\alpha}p^{\beta}\mathds{1}_4\supset -(\mathring{m}_{0}(p^0)^2+\mathring{m}_{2}p^2)\mathds{1}_4\,,
\label{eq1.0}
\end{equation}
where, as of now, $\mathbf{p}^2\equiv p^2$ will be used, for simplicity, which is not to be confused with the four-momentum squared. Furthermore, the isotropic coefficient sets are defined by $\mathring{m}_0\equiv m^{(5)}_{00}$, and $\mathring{m_{2}}\equiv 
m^{(5)}_{11}=m^{(5)}_{22}=m^{(5)}_{33}$. The remaining coefficients do not contribute to the isotropic subsets.

\subsection{Isotropic $\boldsymbol{\mathring{m}_0}$ coefficient}

First of all, we delve into the first isotropic sector, i.e., $\mathring{m}_2=0$. The modified Dirac operator $(\slashed{p}-m_{\psi}+\widehat{\mathcal{Q}})$ in momentum space is consulted, with $\widehat{\mathcal{Q}}$ of Eq.~\eqref{eq1.0} restricted to $\mathring{m}_0$. The existence of nontrivial spinor solutions requires that its determinant vanish, which implies a characteristic equation for $p_0$. Solving the latter provides the possible fermion energies. In total, there are eight dispersion relations for the isotropic coefficient $\mathring{m}_{0}$. Four of them are spurious, i.e., they have no consistent Lorentz-invariant limit; see Ref.~\cite{Marco3}. They are of the form
\begin{align}
E_{\mathring{m}_0}^{(s)\pm}&=\pm\frac{1}{\sqrt{2}}\sqrt{\frac{1}{\mathring{m}_0^2}-\frac{2m_{\psi}}{\mathring{m}_0}+\frac{\sqrt{1-4m_{\psi}\mathring{m}_0-4\mathring{m}_0^2p^2}}{\mathring{m}_0^2}}
\nonumber \\
&=\mp\frac{1}{\mathring{m}_0}\pm m_{\psi} \pm (2m_{\psi}^2+p^2)\frac{\mathring{m}_0}{2}+\dots\,.
\label{eq1.1}
\end{align}
These spurious dispersion relations correspond to Planck scale effects and are singular for $\mathring{m}_0\rightarrow 0$.
Since we are interested in energy regimes way below the Planck scale, the corresponding modes are not excited, whereupon they are disregarded. On the other hand, the remaining four dispersion relations are modified versions of the Lorentz-invariant ones. There are two positive- and two negative-energy dispersion relations and each is spin-degenerate. We only quote the particle dispersion relation:
\begin{align}
E_{\mathring{m}_0}&=\frac{1}{\sqrt{2}}\sqrt{\frac{1}{\mathring{m}_0^2}-\frac{2m_{\psi}}{\mathring{m}_0}-\frac{\sqrt{1-4m_{\psi}\mathring{m}_0-4\mathring{m}_0^2p^2}}{\mathring{m}_0^2}}
\nonumber\\
&=(m_{\psi}\mathring{m}_0+1) E_0(p)+\dots\,, \\[1ex]
E_0(p)&=\sqrt{p^2+m_{\psi}^2}\,,
\label{eq1.2}
\end{align}
with the first-order expansion conveniently expressed in terms of the standard massive dispersion relation $E_0(p)$.

We are now ready to employ Eq.~\eqref{eq1.2} in the energy balance equation $\Delta E=0$, with $\Delta E$ of Eq.~\eqref{ebeq}, to find the polar angle $\mathcal{\theta}$. The exact expression for $\cos \mathcal{\theta}$ is known, but it is involved and intransparent. Hence, the result expanded at first order in $\mathring{m}_0$ is presented:
\begin{equation}
\cos \theta=\frac{1}{q}\Big[E_0(q)+m_{\psi}\mathring{m}_0(k-E_0(q))\Big]\,,
\label{eq1.3}
\end{equation}
where we defined $|\mathbf{q}|\equiv q$.

The magnitude of the photon momentum has an upper limit $k_{\mathrm{max}}$ corresponding to the maximum value of $\cos \theta$. Considering this in Eq.~(\ref{eq1.3}), we find this maximum value:
\begin{align}
k_{\mathrm{max}}&=\frac{q-E_0(q)}{ m_{\psi} \mathring{m}_0}+E_0(q)\,.
\label{eq1.4}
\end{align}
The vacuum Cherenkov process takes place for positive $k$, where a minimum nonzero initial-fermion energy follows from $k_{\mathrm{max}}=0$. Thus, the process only occurs for the initial-fermion momentum lying above a certain threshold. From Eq.~\eqref{eq1.4}, the latter is derived as $q_{\mathrm{th}}\approx \sqrt{m_{\psi}/(2\mathring{m}_0)}$. A numerical treatment reveals that the expression found is approximate and subject to change when higher-order contributions in the expansions with respect to $\mathring{m}_0$ are taken into account. Then,
\begin{equation}
q_{\mathrm{th}}=\sqrt{\frac{m_{\psi}}{3\mathring{m}_0}}+\dots\,,
\label{eq1.5}
\end{equation}
i.e., the global factor changes. Equation~\eqref{eq1.5} tells us that only $\mathring{m}_0>0$ is allowed for the process to occur. Moreover, $q_{\mathrm{th}}$ diverges for $\mathring{m}_0\rightarrow 0$, as expected, since vacuum Cherenkov radiation is energetically forbidden in the Lorentz-invariant limit.

\begin{figure}
\centering
\includegraphics[scale=0.27]{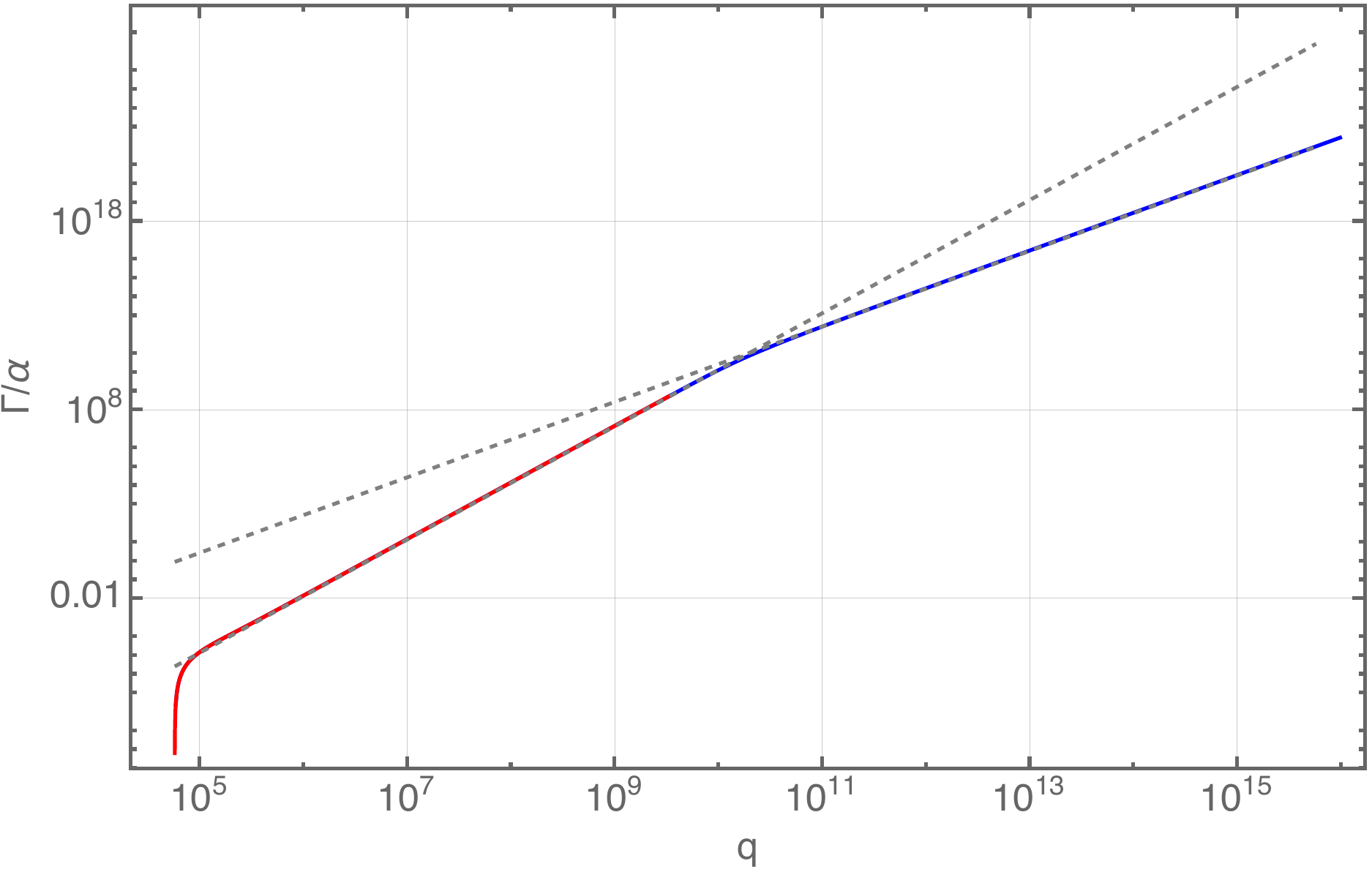}
\caption{Double-logarithmic plot of the decay rate $\Gamma/\alpha$ of vacuum Cherenkov radiation as a function of the initial-fermion momentum $q$ for the isotropic coefficients $\mathring{m}_0$ and $\mathring{m}_2$, represented by solid red and blue lines, respectively. The asymptotic behaviors of Eqs.~\eqref{eq:asymptote-m0}, \eqref{eq:asymptotes-m2} are illustrated by gray, dashed straight lines. We choose $m_{\psi}=\unit[1.00]{GeV}$, $\mathring{m}_0=\unit[10^{-10}]{GeV^{-1}}$, and $\mathring{m}_2=\unit[10^{-10}]{GeV^{-1}}$ such that $q_{\mathrm{th}}\approx \unit[5.77\times 10^4]{GeV}$.}
\label{f1.1}
\end{figure}

To compute the decay rate based on Eq.~(\ref{eq:decay-rate}), we still have to find the amplitude squared and the space phase factor. The sum over the spinor projectors is taken from Eq.~\eqref{eq:sum-spinor-projetors-m}. The matrix element squared depends on $q$, $k$, $m_{\psi}$, and the isotropic coefficient $\mathring{m}_0$, since any dependence on $\theta$ is eliminated according to Eq.~(\ref{eq1.3}). The phase space factor is computed by differentiating the energy balance equation~\eqref{ebeq} for $\theta$:
\begin{subequations}
\begin{align}
\frac{\partial\Delta E_{\mathring{m}_0}}{\partial\theta}&=-(1+m_{\psi}\mathring{m}_0)^2\frac{k q \sin \theta}{E_{\mathring{m}_0}({\bf q}
-{\bf k})}\,, \nonumber\\[1ex]
\Pi(k)&=\frac{k\sin \theta }{E_{\mathring{m}_0}({\bf q}-{\bf k})}\left| \frac{\partial\Delta E_{\mathring{m}_0}}{\partial \mathcal{\theta}} \right|^{-1}\Big|_{\theta=\theta_0} \notag \\
&=\frac{1}{q(1+m_{\psi}\mathring{m}_0)^2}\,,
\end{align}
\end{subequations}
where the latter is independent of the polar angle. Finally, Fig.~\ref{f1.1} presents the numerical result for the decay rate in a double-logarithmic plot. 
As expected, the decay rate goes to zero when the initial-fermion momentum $q$ approaches the threshold. When the fermion momentum is much larger than the threshold, the decay rate exhibits a polynomial dependence of $q$. By analytical means, we find the asymptotic expression for $q\gg q_{\mathrm{th}}$:
\begin{equation}
\label{eq:asymptote-m0}
\Gamma^{\text{(high)}}_{\mathring{m}_0}\sim \frac{27}{20}\alpha\mathring{m}_0^2q^3\,.
\end{equation}
The asymptotic behavior starts at larger values as compared to decay rates computed for isotropic minimal fermion coefficients~\cite{Marco,Schreck:2017egi}. This is also expected, since contributions due to nonminimal coefficients are suppressed at low energies. Note that there is a maximum momentum
\begin{equation}
\label{eq:maximum-momentum-m0}
q_{\mathrm{max}}=\frac{\sqrt{1-4\mathring{m}_0m_{\psi}}}{2\mathring{m}_0}\,,
\end{equation}
beyond which the process ceases to take place. After all, the exact dispersion relation of Eq.~\eqref{eq1.2} takes complex values for $q>q_{\mathrm{max}}$. To the best of our knowledge, such a behavior has not been observed in any minimal setting and it is a consequence of the additional time derivatives present. Nevertheless, as long as $q\in [q_{\mathrm{th}},q_{\mathrm{max}}]$, the decay rate looks well-behaved and there is no indication for unphysical effects. The forthcoming results will corroborate this statement.

\subsection{Isotropic $\boldsymbol{\mathring{m}_2}$ coefficient}

Now we discuss the results for the second independent set of isotropic coefficients such that $\mathring{m}_0=0$. Since the latter are not contracted with additional time derivatives, there are four modified dispersion relations, and none of these is spurious. The fermion energy is given by
\begin{equation}
E_{\mathring{m}_2}=\sqrt{m_{\psi}^2 +(1+ 2 m_{\psi} \mathring{m}_2) p^2 +  \mathring{m}_2^2 p^4}\,.
\label{eq1.6}
\end{equation}
Now, we can find the energy balance equation $\Delta E=0$ based on Eq.~\eqref{ebeq} and solve the latter for the polar angle~$\theta$. The results are involved, but can be computed exactly:
\begin{subequations}
\label{eq1.7}
\begin{align}
\cos \theta&=\frac{qf-\Xi(q)}{8 k \mathring{m}_2^2 q^2}\,, \\[1ex]
\Xi(q)&=\Big\{-8k\mathring{m}_2^2q^2 \left[ k(f-2-2k^2 \mathring{m}_2^2)+4 E_{\mathring{m}_2}(q)\right] \notag \\
&\phantom{{}={}}\quad+q^2f^2\Big\}^{1/2}\,,
\end{align}
where
\begin{align}
f&=f(m_{\psi},\mathring{m}_2,q) \notag \\
&\equiv 4 k^2 \mathring{m}_2^2+4 m_{\psi}\mathring{m}_2+4 \mathring{m}_2^2 q^2+2\,.
\end{align}
\end{subequations}
Unlike for the isotropic coefficient $\mathring{m}_0$ of the previous section, it is possible to solve the energy balance equation for $\theta$ at all orders in Lorentz violation.
The maximum value for the photon momentum can be directly obtained from Eq.~(\ref{eq1.7}) and it reads
\begin{subequations}
\label{eq1.8}
\begin{align}
k_{\mathrm{max}}&=\frac{\sqrt[3]{\sqrt{g^2+32 h^3}+g}}{3 \sqrt[3]{2}} \notag \\
&\phantom{{}={}}-\frac{2 \sqrt[3]{2} h}{3 \sqrt[3]{\sqrt{g^2+32 h^3}+g}}+\frac{4 q}{3}\,,
\end{align}
with
\begin{align}
g=g(m_{\psi},\mathring{m}_2,q)&\equiv -\frac{54 E_{\mathring{m}_2}(q)}{\mathring{m}_2^2}+\frac{36 m_{\psi} q}{\mathring{m}_2} \notag \\
&\phantom{{}={}}+\frac{54 q}{\mathring{m}_2^2}+20 q^3\,, \displaybreak[0]\\[1ex]
h=h(m_{\psi},\mathring{m}_2,q)&\equiv 3 \frac{m_{\psi}}{\mathring{m}_2}+ q^2\,.
\end{align}
\end{subequations}
In this case, the threshold momentum has a behavior analogous to that of Eq.~\eqref{eq1.5}:
\begin{equation}
q_{\mathrm{th}}=\sqrt{\frac{m_{\psi}}{3\mathring{m}_2}}+\dots\,.
\label{mqth}
\end{equation}
As before, the process occurs for positive values of the isotropic coefficient only. The amplitude squared is a function of $q$, $k$, $m_{\psi}$, and the isotropic coefficient $\mathring{m}_2$. The space phase factor is obtained as before and at first order in $\mathring{m}_2$ it reads
\begin{equation}
\Pi(k)=\frac{1}{q}(1-2m_{\psi}\mathring{m}_2)\,.
\label{eq1.9}
\end{equation}
The decay rate for this case is computed numerically and is presented in Fig.~\ref{f1.1}, too. In fact, the curves for $\mathring{m}_0$ considered previously and $\mathring{m}_2$ practically lie on top of each other for $q\in [q_{\mathrm{th}},q_{\mathrm{max}}]$ with $q_{\mathrm{max}}$ for the $\mathring{m}_0$ sector given by Eq.~\eqref{eq:maximum-momentum-m0}. Unlike for the first isotropic sector of $\hat{m}$, the exact dispersion relation of Eq.~\eqref{eq1.6} does not take complex values, which makes the decay rate for $\mathring{m}_2$ continue beyond $q_{\mathrm{max}}$. Interestingly, the behavior of $\Gamma_{\mathring{m}_2}$ for $q\gg q_{\mathrm{th}}$ is characterized by two different asymptotic regimes. For intermediate and high energies, respectively, we find
\begin{subequations}
\label{eq:asymptotes-m2}
\begin{align}
\Gamma_{\mathring{m}_2}^{\text{(inter)}}&\sim \frac{27}{20}\alpha\mathring{m}_2^2q^3\,, \displaybreak[0]\\[2ex]
\label{eq:asymptotes-m2-2}
\Gamma_{\mathring{m}_2}^{\text{(high)}}&\sim \frac{8}{3}\alpha\mathring{m}_2q^2\,,
\end{align}
\end{subequations}
where the intermediate behavior corresponds to the asymptotic form for $\mathring{m}_0$, found in Eq.~\eqref{eq:asymptote-m0}. It is suppressed quadratically in Lorentz violation, but rises more quickly in $q$ than does the high-energy behavior. On the contrary, the latter is linear in Lorentz violation. Comparably, the decay rate in the case of timelike MCFJ theory for nonzero photon mass was found to exhibit two asymptotic regimes, as well \cite{Colladay}.

Now, since the curves for the decay rates for $\mathring{m}_0$ and $\mathring{m}_2$ practically lie on top of each other for $q\in [q_{\mathrm{th}},q_{\mathrm{max}}]$, the additional time derivatives in the first isotropic sector do not seem to impact the dynamics of the process significantly, at least for $q\leq q_{\mathrm{max}}$. Their only consequence is that the process ceases for $q>q_{\mathrm{max}}$, since the fermion energy then becomes imaginary. We therefore conclude that our treatment of $\mathring{m}_0$ is warranted, as long as $q\leq q_{\mathrm{max}}$.

\section{Dimension-5 CPT-odd operator}
\label{s4}

Having gained some understanding of the isotropic parts of $\hat{m}$, let us now dedicate ourselves to the nonminimal dim-5 coefficients $a^{(5)}_{\alpha\beta\mu} $. These are CPT-odd and are contained in the term
\begin{equation}
-\frac{1}{2}(a^{(5)})^{\mu\alpha\beta}\overline{\psi}\gamma_{\mu}\mathrm{i}D_{(\alpha}\mathrm{i}D_{\beta)}\psi+\text{H.c.}\subset \mathcal{L}^{(5)}_{\psi D}\,,
\end{equation}
of Tab.~I in Ref.~\cite{Zonghao}. The operator of Eq.~\eqref{eq:Qhat-a} in momentum space is given by
\begin{equation}
\widehat{\mathcal{Q}}=\slashed{\hat{a}}\supset -(a^{(5)})^{\mu\alpha\beta}p_{\alpha}p_{\beta}\gamma_{\mu}\,,
\end{equation}
and we identify two independent isotropic subsets of coefficients~\cite{Mewes}:
\begin{align}
\slashed{\hat{a}}&\supset(a^{(5)})^{000}p^{0}p^{0}\gamma^{0}+(a^{(5)})^{0jk}p^{j}p^{k}\gamma^{0} \notag \\
&= \mathring{a}^{(5)}_0 p^{0}p^{0}\gamma^{0}+\mathring{a}^{(5)}_2 p^{k}p^{k}\gamma^{0} \nonumber\\
& = \mathring{a}_0 p^{0}p^{0}\begin{pmatrix}
\mathds{1}_2& 0 \\
0 & -\mathds{1}_2
\end{pmatrix}+\mathring{a}_2 \begin{pmatrix}
p^k p^k\mathds{1}_2 & 0 \\
0 & -p^k p^k\mathds{1}_2 \\
\end{pmatrix} \notag \\
&=\begin{pmatrix}
[\mathring{a}_0 (p^{0})^2+\mathring{a}_2p^2]\mathds{1}_2 & 0 \\
0 & -[\mathring{a}_0 (p^{0})^2+\mathring{a}_2p^2]\mathds{1}_2
\end{pmatrix}\,,
\label{eq2.1b}
\end{align} 
where $(a^{(5)})^{000}\equiv \mathring{a}_0$ and $(a^{(5)})^{0ii}=(a^{(5)})^{i0i}=(a^{(5)})^{ii0}\equiv 3\mathring{a}_2$ for the free spatial index $i$.

\subsection{Isotropic $\boldsymbol{\mathring{a}_0}$ coefficient}

\begin{figure}
\centering
\includegraphics[scale=0.27]{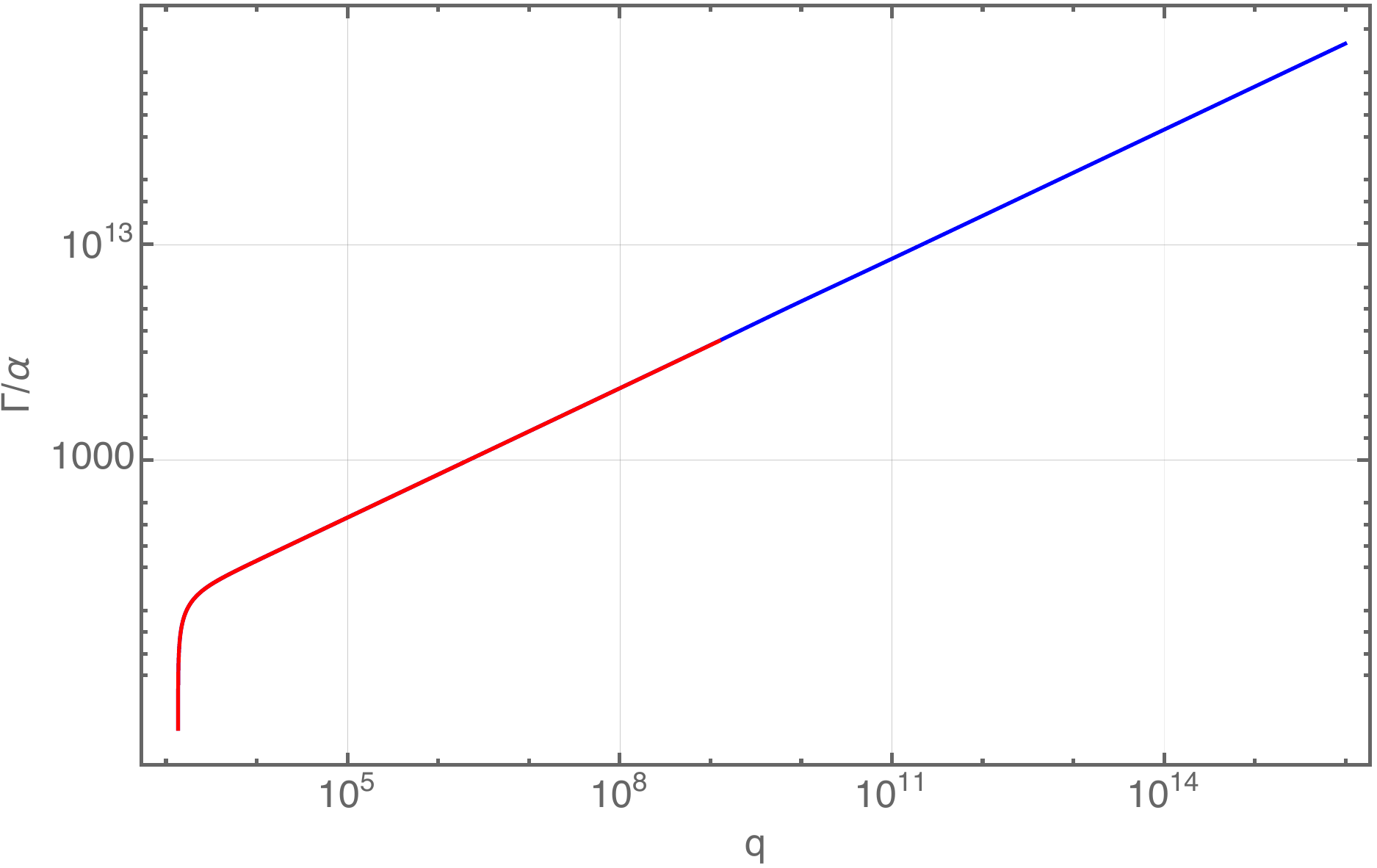}
\caption{The same as Fig.~\protect\ref{f1.1} for the coefficients $\mathring{a}_0$ and $\mathring{a}_2$, respectively. We choose $m_{\psi}=\unit[1.00]{GeV}$, $\mathring{a}_0=\unit[10^{-10}]{GeV^{-1}}$, and $\mathring{a}_2=\unit[10^{-10}]{GeV^{-1}}$ such that $q_{\mathrm{th}}\approx \unit[1360]{GeV}$.}
\label{f1.2}
\end{figure}

Here, we intend to look into the first isotropic case, i.e., $\mathring{a}_2=0$. As before, to calculate the dispersion relations, we require that the determinant of the operator $(\slashed{p}-m_{\psi}+\widehat{\mathcal{Q}})$, with Eq.~\eqref{eq2.1b} restricted to $\mathring{a}_0$, be zero and solve the characteristic equation for $p_0$. For the first isotropic piece $\mathring{a}_0$, we are led to four spurious dispersion relations. An expansion for small $\mathring{a}_0$ shows their divergent behaviors in the limit $\mathring{a}_0\rightarrow 0$:
\begin{align}
\label{eq2.2}
E_{\mathring{a}_0}^{(s)\pm}&=\pm\frac{\sqrt{1-4\mathring{a}_0 E_0(p)}+1}{2 \mathring{a}_0} \notag \\
&=\pm\frac{1}{\mathring{a}_0}\mp E_0(p)+\dots\,.
\end{align}
Again, the latter dispersion relations describe Planck scale effects and are uninteresting for $E\ll E_{\mathrm{Pl}}$. There are two nonspurious positive- and negative-energy dispersion relations, respectively, which are spin-degenerate. In particular, the fermion energy is given by
\begin{align}
E_{\mathring{a}_0}&=\frac{1-\sqrt{1-4\mathring{a}_0E_0(p)}}{2\mathring{a}_0}\notag \\
&=E_0(p)+\mathring{a}_0 E_0^2(p)+\dots\,.
\label{eq2.3}
\end{align}
By resorting to Eq.~(\ref{eq2.3}), the energy balance equation $\Delta E=0$ with $\Delta E$ of Eq.~\eqref{ebeq} is evaluated, whereupon it provides the polar angle at first order in $\mathring{a}_0$:
\begin{equation}
\cos \theta=\frac{1}{q}\left[E_0(q)-\mathring{a}_0 \left(k^2-3 k E_0(q)+2 E^2_0(q)\right)\right]\,.
\label{eq2.4}
\end{equation}
The corresponding maximum value for the photon momentum is obtained from Eq.~\eqref{eq2.4}. At leading order in the isotropic coefficient it reads
\begin{equation}
k_{\mathrm{max}}=\frac{3}{2} E_0(q)-\frac{\sqrt{E_0(q)-q}}{ \sqrt{\mathring{a}_0}}\,.
\label{eq2.5}
\end{equation}
This time it is challenging to determine an analytical expression for the threshold momentum. It is possible to find an approximation, if we consider $q\gg m_{\psi}$: $q_{\mathrm{th}}\approx [2m_{\psi}^2/(9\mathring{a}_0)]^{1/3}$. Then, the process is not allowed for $\mathring{a}_0<0$ and the threshold diverges for $\mathring{a}_0\rightarrow 0$, as expected. 
A numerical treatment reveals that the true threshold is quite close to this value, but reads instead
\begin{equation}
    q_{\mathrm{th}}=\sqrt[3]{\frac{m_{\psi}^2}{4\mathring{a}_0}}+\dots\,.
    \label{aqth}
\end{equation}
As before, the amplitude squared is a function of $k$, $q$, $m_{\psi}$, and $\mathring{a}_0$ after eliminating $\theta$ based on Eq.~\eqref{eq2.4}. The modified spinor solutions of App.~\ref{apA} are employed to compute the sum over spinor bilinears:
\begin{equation}
\label{eq:sum-spinor-projetors-a0}
\sum_s \Lambda^{(s)}_{ab}(q)=\frac{[q^{\alpha}\gamma_{\alpha}+m_{\psi}\mathds{1}_4-\gamma^0\mathring{a}_0(q^0)^2]_{ab}|_{q^0=E_{\mathring{a}_0}(q)}}{\mathring{a}_0E_{\mathring{a}_0}^{(s)+}}\,,
\end{equation}
with the spurious positive-energy dispersion relation of Eq.~\eqref{eq2.2}.
Furthermore, the space phase factor is
\begin{equation}
\Pi(k)=\frac{1}{q}\left[1-3\mathring{a}_0 (E_0(q)-k)\right]\,.
\label{eq2.6}
\end{equation}
The decay rate is obtained numerically and it is presented in the double-logarithmic plot of Fig.~\ref{f1.2}.
As for $\mathring{m}_{0,2}$, the decay rate tends to zero when the momentum $q$ approaches the threshold. The asymptotic behavior starts at lower values of $q$, as compared to the curves for $\mathring{m}_{0,2}$ in Fig.~\ref{f1.1}. In particular, we find
\begin{equation}
\label{eq:decay-rate-asymptotic-a0}
\Gamma_{\mathring{a}_0}^{\text{(high)}}\sim \frac{25}{12}\alpha\mathring{a}_0q^2\,,
\end{equation}
which resembles that of Eq.~\eqref{eq:asymptotes-m2-2}. Similarly to the behavior observed for $\mathring{m}_0$, Eq.~\eqref{eq2.2} takes complex values for $q>\tilde{q}_{\mathrm{max}}$ with
\begin{equation}
\label{eq:qmax-a0}
\tilde{q}_{\mathrm{max}}=\frac{\sqrt{1-16\mathring{a}_0^2m_{\psi}^2}}{4\mathring{a}_0}\,.
\end{equation}
Consequently, vacuum Cherenkov radiation is impossible for momenta larger than $\tilde{q}_{\mathrm{max}}$. This characteristic is presumably again a consequence of the additional time derivatives. Nevertheless, as long as $q\in [q_{\mathrm{th}},\tilde{q}_{\mathrm{max}}]$, the decay rate is well-behaved and does not exhibit any unphysical properties.

\subsection{Isotropic $\boldsymbol{\mathring{a}_2}$ coefficient}

Finally, we delve into the second isotropic part of the dim-5 $a$ coefficients: $\mathring{a}_2$ from Eq.~\eqref{eq2.1b}, i.e., $\mathring{a}_0=0$. There are no spurious dispersion relations for this case and the two positive-energy dispersion relations are spin-degenerate, as before. At first order in $\mathring{a}_2$, the fermion energy is given by:
\begin{equation}
E_{\mathring{a}_2}=E_0(p)+\mathring{a}_2 p^2+\dots\,.
\label{eq2.7}
\end{equation}
Note that the exact dispersion relation is quite involved, which renders any follow-up evaluation of the decay rate highly challenging. Even for a numerical computation, we must rely on certain analytical results. Therefore, we will continue with Eq.~\eqref{eq2.3} expanded at first order in Lorentz violation. However, one may then question the interpretation of forthcoming higher-order results in Lorentz violation. Interestingly, by taking the choice $(a^{(5)})^{0ii}\equiv \mathring{a}_2$, which is a subset of the second isotropic sector, as defined under Eq.~\eqref{eq2.1b}, the first-order dispersion relation of Eq.~\eqref{eq2.7} turns out to be exact. When the current analysis is based on this subset of coefficients, all contributions at higher orders in Lorentz violation make sense. Thus, we will proceed along these lines.

With Eq.~\eqref{eq2.7} at our disposal, it is possible to solve the energy balance equation $\Delta E=0$ for $\theta$:
\begin{equation}
\cos \theta=\frac{1}{q}\left[E_0(q)-\mathring{a}_2 (k^2-3 k E_0(q)+2E^2_0(q))\right]\,.
\label{eq2.8}
\end{equation}
As done previously, by requiring that $\cos \theta$ assume its maximum value, we find the maximum value for the photon momentum:
\begin{equation}
k_{\mathrm{max}}=\frac{3}{2}E_0(q)-\frac{\sqrt{E_0(q)-q}}{\sqrt{\mathring{a}_2}}\,.
\label{eq2.9}
\end{equation}
Note that Eqs.~\eqref{eq2.8}, \eqref{eq2.9} correspond to Eqs.~\eqref{eq2.4} and \eqref{eq2.5}, respectively, with only the coefficients interchanged. Here, the threshold energy can be computed analytically:
\begin{equation}
q_{\mathrm{th}}=\sqrt[3]{\frac{m_{\psi}^2}{4\mathring{a}_2}}+\dots\,,
\label{eq:threshold-a2}
\end{equation}
cf.~Eq.~\eqref{aqth}.
To perform the numerical integration over the phase space variables, we write the amplitude squared as a function of $k$, $q$, $m_{\psi}$, and $\mathring{a}_2$. The sum over the spinor bilinears is found to be
\begin{align}
\label{eq:sum-spinor-projetors-a0}
\sum_s \Lambda^{(s)}_{ab}(q)&=\left(1+\mathring{a}_2\frac{q^2}{E_0(q)}\right) \notag \\
&\phantom{{}={}}\times[q^{\alpha}\gamma_{\alpha}+m_{\psi}\mathds{1}_4-\gamma^0\mathring{a}_2q^2]_{ab}|_{q^0=E_{\mathring{a}_2}(q)}\,,
\end{align}
and the phase space factor is expressed as follows:
\begin{equation}
\Pi(k)=\frac{1}{q}\left[1-\mathring{a}_2\left(3(E_0(q)-k)-\frac{m_{\psi}^2}{E_0(q)-k}\right)\right]\,.
\label{eq2.10}
\end{equation}
Figure~\ref{f1.2} presents the decay rate obtained from integrating the amplitude squared over the phase space numerically. As for $\mathring{m}_{0,2}$, the curves for $\mathring{a}_{0,2}$ are congruent as long as $q\leq \tilde{q}_{\mathrm{max}}$ with the maximum momentum $\tilde{q}_{\mathrm{max}}$ for $\mathring{a}_0$ given by Eq.~\eqref{eq:qmax-a0}. On the contrary to $\mathring{a}_0$, the dispersion relation for $\mathring{a}_2$ does not become complex. Hence, $\Gamma_{\mathring{a}_2}$ continues beyond $\tilde{q}_{\mathrm{max}}$ and we again discover that there is an intermediate- and a high-energy behavior:
\begin{subequations}
\begin{align}
\label{eq:decay-rate-assymptotic-a2-1}
\Gamma^{\text{(inter)}}_{\mathring{a}_2}\sim \frac{25}{12}\alpha\mathring{a}_2q^2\,, \\[2ex]
\label{eq:decay-rate-assymptotic-a2-2}
\Gamma^{\text{(high)}}_{\mathring{a}_2}\sim \frac{32}{15}\alpha\mathring{a}_2q^2\,,
\end{align}
\end{subequations}
cf.~Eq.~\eqref{eq:decay-rate-asymptotic-a0}.
Unlike for $\mathring{m}_2$, both asymptotic behaviors only differ in their global factors. Thus, the decay rate is not suppressed at second order in Lorentz violation for intermediate energies. Note that the slight change in behavior from Eq.~\eqref{eq:decay-rate-assymptotic-a2-1} to \eqref{eq:decay-rate-assymptotic-a2-2} is very subtle and cannot simply be perceived in Fig.~\ref{f1.2}. This is in stark contrast to the markant change of slopes in the double-logarithmic plot for $\mathring{m}_2$; cf.~Fig.~\ref{f1.1}. Therefore, we emphasize the importance of supporting numerical computations by analytical ones, whenever possible.

\section{Radiated-energy rate}
\label{sec:radiated-energy-rate}

Moreover, we also compute the radiated-energy rates for each coefficient set according to Eq.~\eqref{eq:radiated-energy-rate}, which play an essential role for phenomenology. In fact, the resulting curves have quite similar characteristics to those for the decay rates. Since there is nothing new to be gained from these curves, we refrain from stating them here explicitly. However, for completeness, we provide information on the asymptotic regimes of this quantity, since the latter can again be obtained analytically. In generic form, the asymptotic behaviors for the decay rate and the radiated-energy rate, respectively, are expressed as 
\begin{subequations}
\label{eq:generic-formulas-rates}
\begin{align}
\Gamma&=\xi\alpha\mathring{X}^sq^t\,, \\[2ex]
\frac{\mathrm{d}W}{\mathrm{d}t}&=\zeta\alpha\mathring{X}^sq^{t+1}\,,
\end{align}
\end{subequations}
in terms of $\xi,\zeta\in\mathbb{R}_+$ and $s,t\in\mathbb{N}$. For dimensional reasons, the power of the momentum in the radiated-energy rate is larger by 1 as compared to the momentum dependence of the decay rate. Technically, the intermediate-energy behavior is obtained by expanding the integrand in both the LV coefficient and the fermion mass at leading order. To extract the high-energy behavior, the integrand is expanded in the inverse fermion momentum $1/q$. Values for the parameters in Eq.~\eqref{eq:generic-formulas-rates} for the different sectors investigated are stated in Tab.~\ref{tab:parameters-asymptotic-regimes}.
\begin{table}[t]
\centering
\begin{tabular}{cccccc}
\toprule
Coefficient & $\xi$ & $\zeta$ & $s$ & $t$ \\
\midrule
$\mathring{m}_0$ & 27/20 & 29/60 & 2 & 3 \\
$\mathring{m}_2$ & 27/20 & 29/60 & 2 & 3 \\
                 & 8/3 & 2 & 1 & 2 \\
$\mathring{a}_0$ & 25/12 & 4/5 & 1 & 2 & \\
$\mathring{a}_2$ & 25/12 & 4/5 & 1 & 2 & \\
                 & 32/15 & 4/3 & 1 & 2 & \\
\bottomrule
\end{tabular}
\caption{Parameters for asymptotic behaviors of decay and radiated-energy rates according to Eq.~\eqref{eq:generic-formulas-rates}.}
\label{tab:parameters-asymptotic-regimes}
\end{table}%

Having the asymptotic behaviors of the radiated-energy rates at our disposal, we can compute the particle energy $E$ as a function of time $t$. It follows from the first-order ordinary differential equation
\begin{equation}
\label{eq:energy-loss}
\dot{E}=-\frac{\mathrm{d}W}{\mathrm{d}t}\,,
\end{equation}
with the initial condition $E(t=0)=E_0$, where $E_0>E_{\mathrm{th}}$ is the energy of the particle right after it is produced by an astrophysical source not specified further. Since $E_{\mathrm{th}}\gg m_{\psi}$, it also holds that $E\simeq q$, where $q$ is the particle momentum used previously. In the asymptotic regimes, Eq.~\eqref{eq:energy-loss} can be solved analytically via basic methods. For example, for $\mathring{m}_0$ we have that
\begin{equation}
E^{\text{(high)}}_{\mathring{m}_0}(t)=\left(\frac{1}{E_0^3}+\frac{3\zeta}{\hbar}\alpha\mathring{m}_0^2t\right)^{-1/3}\,,
\end{equation}
where $\hbar$ has been reinstated. For $\mathring{m}_2$, the behavior in the intermediate-energy regime is analogous with $\mathring{m}_0$ replaced by $\mathring{m}_2$. The high-energy behavior is substantially different. In total, we arrive at
\begin{subequations}
\begin{align}
E_{\mathring{m}_2}^{\text{(inter)}}&=\left(\frac{1}{E_0^3}+\frac{3\zeta}{\hbar}\alpha\mathring{m}_2^2t\right)^{-1/3}\,, \displaybreak[0]\\[2ex]
\label{eq:energy-time-m2}
E_{\mathring{m}_2}^{\text{(high)}}&=\left(\frac{1}{E_0^2}+\frac{2\zeta}{\hbar}\alpha\mathring{m}_2t\right)^{-1/2}\,.
\end{align}
\end{subequations}
For both $\mathring{a}_0$ and $\mathring{a}_2$, the behaviors are described by Eq.~\eqref{eq:energy-time-m2}, where the SME coefficient and $\zeta$ must be replaced suitably.

Since we employed the asymptotic behaviors, the initial energy $E_0$ should not be chosen too close to the threshold. A reasonable choice is $E_0=2E_{\text{th}}$ with the appropriate threshold energy of Eqs.~\eqref{eq1.5}, \eqref{mqth}, \eqref{aqth}, and \eqref{eq:threshold-a2}, respectively. When taking the proper values for $\zeta$ from Tab.~\ref{tab:parameters-asymptotic-regimes}, it is possible to demonstrate that the radiating fermion already loses a significant amount of its surplus energy after fractions of seconds.

\section{Phenomenology and constraints}
\label{s5}

We are now ready to use data from UHECRs to set bounds on the isotropic nonminimal dim-5 coefficients of the previous section.
The threshold energy of the vacuum Cherenkov process is written in the general form
\begin{equation}
\label{eq:threshold-generic}
E_{\mathrm{th}}=\frac{\rho m_{\psi}^{\lambda}}{\mathring{X}^{\sigma}}\,,
\end{equation}
where $m_{\psi}$ is the fermion mass, $\rho$ and $\sigma$ are dimensionless parameters, and $\mathring{X}$ is an isotropic dim-5 coefficient; see Tab.~\ref{tab:parameters-threshold} for explicit values for the coefficients $\mathring{m}_{0,2}$ and $\mathring{a}_{0,2}$ studied. When an UHECR event is observed on Earth, its energy should be smaller than the threshold for vacuum Cherenkov radiation. Otherwise, as we argued in Sec.~\ref{sec:radiated-energy-rate}, it would have radiated away its surplus energy on a time scale significantly smaller than typical ones for propagation over astrophysical distances.

The condition $E<E_{\mathrm{th}}$ then leads to the inequality
\begin{equation}
\label{eq:constraint-generic}
\mathring{X}<\left(\frac{\rho m_{\psi}^{\lambda}}{E}\right)^{1/\sigma}\,.
\end{equation}
To obtain a constraint on the coefficient $\mathring{X}$ at the $2\sigma$ level, we add twice the uncertainty related
to the experimental error $\Delta E$ of the energy measured:
\begin{equation}
\label{eq:formula-constraints}
\mathring{X}<\left(\frac{\rho m_{\psi}^{\lambda}}{E}\right)^{1/\sigma}+2\Delta E\left|\frac{\partial}{\partial E}\left(\frac{\rho m_{\psi}^{\lambda}}{E}\right)^{1/\sigma}\right|\,.
\end{equation}
Now we explicitly compute bounds on Lorentz violation based on the energy of a primary cosmic ray detected by the Pierre-Auger observatory \cite{CosmicRays}. We choose the event $737165$ from Tab.~I of this reference, whose energy is slightly increased to $E=\unit[212\times 10^{18}]{eV}$. The latter value is based on assuming a hadronic primary instead of a photon primary; see Tab.~1 of Ref.~\cite{Klinkhamer3}. The energy uncertainty $\Delta E/E$ amounts to 25\%. The composition of the primary hadron is unknown. We repeat the analysis performed for the coefficients of the minimal SME \cite{Marco,Schreck:2017egi}, where an iron nucleus with $N = 56$ nucleons was taken as a conservative choice. Then, the energy of each nucleon is $E/56$.

At this point, we have two possibilities of proceeding. The first is to assume that the Cherenkov photons are emitted by a proton, treated as a pointlike Dirac particle. Doing so enables us to constrain nonminimal Lorentz violation in protons. Alternatively, the internal structure of the nucleons is taken into account. In this case, the Cherenkov photons are emitted from the real up or down-type quarks in the nucleons. This approach then enables us to obtain constraints on nonminimal Lorentz violation in quarks. Note that the structure of the operator $\hat{m}$ is compatible with \textit{U}(1) charge symmetry \cite{ColKost}, such that we can resort to the extended QED of Eq.~\eqref{eq:modified-QED}. Effects of the strong interaction are put aside for this alternative evaluation. The up and down quark masses are taken as $m_{\mathrm{u}}\approx \unit[2.16\times 10^{-3}]{GeV}$ and $m_{\mathrm{d}}\approx \unit[4.70\times 10^{-3}]{GeV}$~\cite{ParticleDataGroup:2024cfk}. Also, these quarks are assumed to carry a fraction $r=0.1$ of the nucleon energy~\cite{quarkfrac}.
\begin{table}
\begin{tabular}{cccc}
\toprule
Coefficient & $\rho$ & $\lambda$ & $\sigma$ \\
\midrule
$\mathring{m}_0$ & $1/\sqrt{3}$ & 1/2 & 1/2 \\
$\mathring{m}_2$ & $1/\sqrt{3}$ & 1/2 & 1/2 \\
$\mathring{a}_0$ & $1/\sqrt[3]{4}$ & 2/3 & 1/3 \\
$\mathring{a}_2$ & $1/\sqrt[3]{4}$ & 2/3 & 1/3 \\
\bottomrule
\end{tabular}
\caption{Parameters of the generic threshold formula of Eq.~\eqref{eq:threshold-generic} for the different isotropic coefficients considered.}
\label{tab:parameters-threshold}
\end{table}

Moreover, in the charged-lepton sector we benefit from the recently completed Large High Altitude Air Shower Observatory (LHAASO) in China, which is a state-of the-art instrument achieving unprecedented surveys of PeV photons from the Crab Nebula~\cite{LHAASO:2021cbz}. These photons are explained via inverse-Compton scattering effects with ultra-high-energy electrons. It is deduced that for the production of a $\unit[1.1]{PeV}$ photon, the parent electron must have had an energy of $\unit[2.3]{PeV}$. The square kilometer array (KM2A), which is used for photon detection, has an energy resolution $\Delta E/E$ of $\leq 20\%$. As proposed in Ref.~\cite{Li:2025uwn}, taking these experimental data as a foundation, nonminimal Lorentz violation in the electron sector can be constrained strictly. Similarly to the quark sector, the nonminimal operator $\hat{m}$ is compatible with $\mathit{U}(1)$ charge symmetry, i.e., the weak interaction is not considered here.

The constraints then follow from Eq.~\eqref{eq:formula-constraints} based on the values of Tab.~\ref{tab:parameters-threshold}. 
The results for the constraints for each of the isotropic coefficients taken into account are presented in Tab.~\ref{tab1}. Although the Lorentz and CPT-violating coefficients can assume positive or negative values, only positive ones are allowed in Eqs.~\eqref{eq1.5}, \eqref{mqth}, and \eqref{aqth}. This implies that the bounds are only one-sided. To obtain two-sided bounds, a complementary process such as photon decay must be explored. Doing so may pose an interesting project that is left for the future.
\begin{table}
\centering
\begin{tabular}{ccc}
 \toprule
 Fermion sector& Isotropic coefficient & Upper Constraint   \\
 \midrule
proton & $\mathring{m}_0$ & $<\unit[1\times 10^{-18}]{GeV^{-1}}$ \\
       & $\mathring{m}_2$ & $<\unit[1\times 10^{-18}]{GeV^{-1}}$ \\
       & $\mathring{a}_0$ & $<\unit[3\times 10^{-28}]{GeV^{-1}}$ \\
       & $\mathring{a}_2$ & $<\unit[3\times 10^{-28}]{GeV^{-1}}$ \\
\midrule
 u quark& $\mathring{m}_0$ & $< \unit[3\times 10^{-18}]{GeV^{-1}}$ \\
         & $\mathring{m}_2$  & $< \unit[3\times 10^{-18}]{GeV^{-1}}$ \\
         & $\mathring{a}_0$ & $<\unit[2\times 10^{-29}]{GeV^{-1}}$   \\
         & $\mathring{a}_2$ & $<\unit[2\times 10^{-29}]{GeV^{-1}}$\\
 \midrule
 d quark& $\mathring{m}_0$ & $< \unit[6\times 10^{-18}]{GeV^{-1}}$ \\
           & $\mathring{m}_2$  & $< \unit[6\times 10^{-18}]{GeV^{-1}}$  \\        
            & $\mathring{a}_0$ & $<\unit[9\times 10^{-29}]{GeV^{-1}}$  \\
            & $\mathring{a}_2$ & $<\unit[9\times 10^{-29}]{GeV^{-1}}$ \\
\midrule
electron & $\mathring{m}_0$ & $<\unit[6\times 10^{-17}]{GeV^{-1}}$ \\
         & $\mathring{m}_2$ & $<\unit[6\times 10^{-17}]{GeV^{-1}}$ \\
         & $\mathring{a}_0$ & $<\unit[1\times 10^{-26}]{GeV^{-1}}$ \\
         & $\mathring{a}_2$ & $<\unit[1\times 10^{-26}]{GeV^{-1}}$ \\
 \bottomrule
\end{tabular}
\caption{Constraints on isotropic controlling coefficients in protons, up and down quarks, and electrons at the $2\sigma$ level based on the event $737165$ detected by the Pierre-Auger observatory, see Refs.~\cite{CosmicRays,Klinkhamer3}, and recent LHAASO measurements~\cite{LHAASO:2021cbz}.}
\label{tab1}
\end{table}

Interestingly, the constraints on $\mathring{m}_{0,2}$ for protons and electrons lie in the same ballpark. The constraints on $\mathring{a}_{0,2}$ for protons are better than the corresponding ones for electrons due to the higher energy of the hadron primary as compared to the parent electron in the inverse Compton process. For $\mathring{m}_{0,2}$, the constraints in quarks are slightly weaker than those for protons, whereas the behavior is opposite for $\mathring{a}_{0,2}$.
According to the data tables~\cite{DataTables}, the dim-5 $a$ coefficients have already been constrained for u and d quarks via Drell-Yan and deep inelastic scattering processes \cite{ZEUS:2022msi,Lunghi:2024ctv}. Most of these bounds are at the $10^{-6}$ -- $\unit[10^{-7}]{GeV^{-1}}$ level. The presented constraints are stronger, but they also rely on more assumptions than do the bounds of Refs.~\cite{ZEUS:2022msi,Lunghi:2024ctv}, which rest upon data collected in controlled environments.

Limits on the dim-5 $a$ coefficients have also been obtained for electrons, which mainly lie within the range $10^{-7}$ -- $\unit[10^{-8}]{GeV^{-1}}$ and result from atomic-physics experiments. Note that there are constraints on $\mathring{a}^{\mathrm{UR}(5)}$ for electrons at the order of magnitude of $\unit[10^{-27}]{GeV}$ \cite{Mewes,Konopka:2002tt,Jacobson:2002ye}, which are comparable to ours. Other than that, constraints on dim-5 $m$ coefficients for electrons and quarks are generically not available, since the form of this operator is incompatible with $\mathit{SU}(2)_L$ and $\mathit{SU}(3)_c$ invariance, i.e., it cannot be deduced from the SME. However, these coefficients can still be considered at the level of a modified QED, which is what we did and how the corresponding bounds of Tab.~\ref{tab1} must be interpreted.

\section{Conclusions}
\label{s6}

This manuscript reports on the intriguing phenomenon of vacuum Cherenkov radiation for a modified QED with isotropic dim-5 SME operators in Dirac fermions. Explicitly, two LV dim-5 operators of different CPT-handedness were considered, which contain the controlling coefficients $m^{(5)}_{\alpha\beta}$ and $a^{(5)}_{\alpha\beta\gamma}$, respectively~\cite{Mewes,Zonghao}. We resorted our study to the isotropic sectors of these coefficient sets, which are governed by $\mathring{m}_{0,2}$ and $\mathring{a}_{0,2}$. Both the kinematics and the dynamics of the vacuum Cherenkov process were evaluated, providing analytical expressions for the threshold energies and numerically evaluated decay rates in terms of the incoming-fermion momentum.

The threshold energies for $\mathring{m}_{0,2}$ and $\mathring{a}_{0,2}$ each were found to be analogous, whereas their functional form differs between $\hat{m}$ and $\hat{a}^{\mu}$. Moreover, apart from their thresholds, the processes for $\mathring{a}_0$ and $\mathring{m}_0$ have maximal momenta beyond which they are no longer allowed. It is interesting to note that the decay rates for $\mathring{m}_2$ and $\mathring{a}_2$, respectively, have two distinct asymptotic regimes. For $\mathring{m}_2$, the polynomial dependence on the momentum differs in these two regimes, which is not the case for $\mathring{a}_2$, though.

Based on an effective modified QED and experimental data on UHECRs and photons, we have been able to constrain isotropic nonminimal Lorentz violation in protons, quarks, and electrons. The sensitivity for these nonminimal coefficients lies at the level of the Planck scale and beyond. By means of the SME, we were able to quantify the absence of Lorentz-violating signals in UHECRs in terms of constraints on SME coefficients. Since any violation of spacetime symmetries are Planck-scale effects, our constraints clearly show that there is no dim-5 isotropic Lorentz violation at the expected energy scales. This poses severe restrictions on models of quantum gravity, since the existence of a quantum regime for gravity is expected to lead to such effects, as discussed briefly at the beginning of the paper.

Let us also refer to the possible scenarios for other dim-5 operators compiled in Ref.~\cite{Zonghao}. We note that within our calculations, all results have been obtained through the dispersion relations for free modified Dirac fermions, which are subject to single-photon emission according to Fig.~\ref{fig0}. Therefore, we can conclude that all dim-5 operators ${\cal L}^{(5)}_{\psi F}$ from Ref.~\cite{Zonghao}, proportional to the electromagnetic field strength tensor $F_{\mu\nu}$, do not contribute to the decay rate of this process at tree-level, as they do not modify the dispersion relations for free fermions.

Now, there are several possibilities of how to extend this work. The first is to look at isotropic sectors of more involved operators such as $\hat{c}^{\mu\nu}$. Second, an analysis can be carried out on anisotropic sectors of $\hat{m}$ and $\hat{a}^{\mu}$ or possibly other operators. Third, we propose to analyze spin-flip processes, which are presumably induced by spin-nondegenerate operators such as $\hat{b}^{\mu}$; see Ref.~\cite{Marco,Schreck:2017egi} for minimal coefficients. Fourth, the present study is to be extended via a broad analysis of experimental data instead of resorting to two single events. These problems pose potentially intriguing projects as well as technical challenges. Their study may provide further insight into the nature of vacuum Cherenkov radiation for nonminimal Lorentz violation.

\section*{Acknowledgments}

The work of A.Yu.\ P. has been partially supported by the CNPq Project No.~303777/2023-0. M.S. is indebted to CNPq Produtividade~310076/2021-8 and CAPES/Finance Code~001.

\appendix

\section{Modified spinor solutions}
\label{apA}

Here, we state the explicit modified spinor solutions of the Dirac equation in momentum space, where the Dirac representation is employed. For the coefficients $\mathring{m}_{0,2}$, the positive-energy spinors are given by
\begin{subequations}
\begin{align}
u^{(1)}&=N_{\mathring{m}_{0,2}}\begin{pmatrix}
1 \\
0 \\
\Xi_{\mathring{m}_{0,2}}\frac{p^3}{p} \\
\Xi_{\mathring{m}_{0,2}}\frac{p^1+\mathrm{i}p^2}{p} \\
\end{pmatrix}\,, \\[0.2cm]
u^{(2)}&=N_{\mathring{m}_{0,2}}\begin{pmatrix}
0 \\
1 \\
\Xi_{\mathring{m}_{0,2}}\frac{p^1-\mathrm{i}p^2}{p} \\
-\Xi_{\mathring{m}_{0,2}}\frac{p^3}{p} \\
\end{pmatrix}\,,
\end{align}
with the normalization factor
\begin{equation}
N_{\mathring{m}_{0,2}}=\sqrt{\frac{2E_{\mathring{m}_{0,2}}p^2}{p^2+\Xi_{\mathring{m}_{0,2}}^2}}
\end{equation}
and
\begin{align}
\Xi_{\mathring{m}_0}&=(1-\mathring{m}_0E_{\mathring{m}_0})E_{\mathring{m}_0}-m_{\psi}\,, \displaybreak[0]\\[1ex]
\Xi_{\mathring{m}_2}&=E_{\mathring{m}_0}-\mathring{m}_2p^2-m_{\psi}\,.
\end{align}
\end{subequations}
Here, $|\mathbf{p}|\equiv p$, for simplicity, where $\mathbf{p}^2\equiv p^2$ should not be confused with the second component of the spatial momentum. The spinors are normalized according to
\begin{equation}
u^{(\alpha)\dagger}u^{(\beta)}=2E_{\mathring{m}_{0,2}}\delta^{\alpha\beta}\,.
\end{equation}
In fact, the spinor solutions for $\mathring{a}_{0,2}$ are very similar in form under the replacements $\mathring{m}_{0,2}\mapsto \mathring{a}_{0,2}$ and $E_{\mathring{m}_{0,2}}\mapsto E_{\mathring{a}_{0,2}}$. As expected, the modified spinors reproduce the well-known standard ones in the Dirac representation for $\mathring{m}_{0,2}\rightarrow 0$ and $\mathring{a}_{0,2}\rightarrow 0$, respectively.

\end{document}